\documentclass[letterpaper, 10 pt, conference]{IEEEtran}
\IEEEoverridecommandlockouts
\usepackage[letterpaper, left=0.75in, right=0.75in, bottom=0.8in, top=0.75in]{geometry}
\usepackage{cite}
\usepackage{lipsum}
\usepackage{booktabs}
\usepackage{amsmath,amssymb,amsfonts}
\usepackage{algorithmic}
\usepackage{graphicx}
\usepackage{textcomp}
\usepackage{xcolor}
\usepackage{graphicx}
\usepackage{subcaption}
\usepackage{multirow}
\usepackage{hyperref}
\usepackage{mwe}
\usepackage{pgfplots}
\def\BibTeX{{\rm B\kern-.05em{\sc i\kern-.025em b}\kern-.08em
    T\kern-.1667em\lower.7ex\hbox{E}\kern-.125emX}}

\newcommand\myfootnote[1]{%
  \begingroup
  \renewcommand\thefootnote{}\footnote{#1}%
  \addtocounter{footnote}{-1}%
  \endgroup
}

\begin{document}

\title{\vspace{0.25in}Neural Network Based Lidar Gesture Recognition for Realtime Robot Teleoperation\\
\thanks{This work was conducted at Defence Research and Development Canada. Prof. Grondin provided subject matter advise to the research team. S. Chamorro was employed as a student researcher when this research was conducted. }
}

\author{Simon Chamorro, Jack Collier, Fran\c{c}ois Grondin\thanks{S. Chamorro and F. Grondin are with the Department of Electrical Engineering and Computer Engineering, Interdisciplinary Institute for Technological Innovation (3IT), 3000 boul. de l'Universit\'e, Universit\'e de Sherbrooke, Sherbrooke, Qu\'ebec (Canada) J1K 0A5, \texttt{\{simon.chamorro,francois.grondin2\}@usherbrooke.ca}.}\thanks{J. Collier is with Defence Research and Development Canada, Suffield Research Centre, 4000 Stn Main, Medicine Hat, Alberta (Canada) T1A 8K6, \texttt{jack.collier@drdc-rddc.gc.ca}.}


\thanks{{\bf \copyright 2021 IEEE. Personal use of this material is permitted. Permission from IEEE must be obtained for all other uses, in any current or future media, including reprinting/republishing this material for advertising or promotional purposes, creating new collective works, for resale or redistribution to servers or lists, or reuse of any copyrighted component of this work in other works.}}


}
\maketitle


\begin{abstract}
We propose a novel low-complexity lidar gesture recognition system for mobile robot control robust to gesture variation. 
Our system uses a modular approach, consisting of a pose estimation module and a gesture classifier.
Pose estimates are predicted from lidar scans using a Convolutional Neural Network trained using an existing stereo-based pose estimation system.  
Gesture classification is accomplished using a Long Short-Term Memory network and uses a sequence of estimated body poses as input to predict a gesture.
Breaking down the pipeline into two modules reduces the dimensionality of the input, which could be lidar scans, stereo imagery, or any other modality from which body keypoints can be extracted, making our system lightweight and suitable for mobile robot control with limited computing power. 
The use of lidar contributes to the robustness of the system, allowing it to operate in most outdoor conditions, to be independent of lighting conditions, and for input to be detected 360 degrees around the robot. 
The lidar-based pose estimator and gesture classifier use data augmentation and automated labeling techniques, requiring a minimal amount of data collection and avoiding the need for manual labeling. 
We report experimental results for each module of our system and demonstrate its effectiveness by testing it in a real-world robot teleoperation setting.
\end{abstract}



\begin{figure}[ht]
    \centering
  \includegraphics[width=0.9\linewidth]{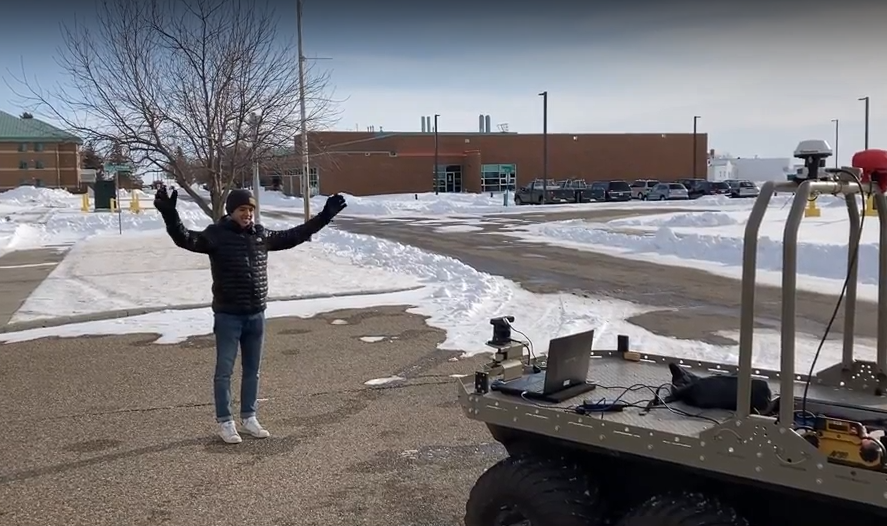}
  \caption{Gesture-based vehicle teleoperation.}
  \label{fig:photo}
\end{figure}

\section{Introduction}

Robotic systems are expected to play an increasingly important role in everyday life whether it be autonomous vehicles, personal assistant robotics, etc. Large-scale adoption of these systems into society is dependant, in part, on their ease of use. human-robot interaction (HRI) mechanisms must be intuitive, robust and efficient without imposing a heavy cognitive burden on the user. An effective HRI interface will work regardless of the environment and without cumbersome specialized training. These HRI principles are especially true in defence applications where the soldier must be aware of their surroundings and be ready to react to any dangers. One potential HRI technology that may help alleviate cognitive burden while being intuitive is gesture-based HRI.

Traditional gesture-based control uses instrumentation such as inertial measurement units (IMUs) or touch-sensitive gloves to detect gestures. These technologies require robust communications infrastructure and specialized equipment to ensure the robot receives commands. Camera-based systems use machine learning techniques to determine gestures but lack robustness to varying lighting conditions, are unusable in the dark, and often have a limited field of view.



The HRI design goals and the deficiencies of current systems listed above, motivated the authors at Defence R\&D Canada to investigate lidar-based gesture recognition to enable soldier-robot teaming. In our intial work \cite{Kealey2020Gesture}, we adapted the pedestrian tracking and classification work in \cite{kenji} to address large gesture classification for robot control. Slice features \cite{slice}, which essentially encode the contour profile of a person, were used to train a model using Adaboost and learn gestures for robot teleoperation. While basic control was observed, the algorithm did not have a temporal component and could only detect static gestures and lacked robustness at moderate distances from the lidar due to point cloud density on a per frame basis. The system also required the area to be pre-mapped using GraphSlam \cite{2011GrisettiG2O} to allow for effective clustering of humans before gesture classification.
\myfootnote{A demo video is available at \url{https://youtu.be/jb-rMVhhTkk}}
In this work, we take a neural-network approach which includes a temporal component allowing us to classify both static and dynamic gestures using a lidar. 
We develop a lidar-based gesture recognition system for soldier-robot teaming. The system is able to operate in all lighting conditions, with no specialized user instrumentation or communications hardware. Furthermore, the operator can control the robot anywhere within the lidar's 360 degree field of view.  The system has been integrated onto a large ground robot allowing for gesture-based teleoperation as well as autonomous leader-follower behaviours.
To the knowledge of the authors, this is a novel approach to gesture and robot control. In particular, the authors have not seen any other system which uses sparse lidar to enable teleoperation in outdoor environments using learned gesture recognition.

The main contributions of the work are as follows:
\begin{enumerate}
    \item A robust and low-complexity large gesture classification system that runs in real-time using an  long  short-term  memory (LSTM) model.
    \item A pose estimation pipeline from lidar that uses a convolutional neural network (CNN) to learn body-pose features.
    \item A gesture-based teleoperation system for a large ground vehicle operating in outdoor environments.
\end{enumerate}


\section{Related Work}
Gesture detection and activity recognition is an active area of research that can benefit to numerous real-life applications \cite{mitra2007gesture}.
For example, activity recognition is used for monitoring purposes in the healthcare, security and surveillance industries \cite{ranasinghe2016review}.
Similarly, gesture recognition is very promising for human-robot interaction (e.g. smart home applications \cite{wan2014gesture}, self-parking for cars \cite{amara2019end} and even robot surgery \cite{van2021gesture}). In defence research, related work \cite{Taylor1} looks at how a soldier interacts with a wheeled robot to accomplish a task. The authors developed a system that uses a 9-axis IMU to perform gesture-based teleoperation. The work is further integrated with speech in \cite{Taylor2} to allow for multi-modal commands. 

Most of the systems introduced above either use wearable sensors to detect human gestures, or focus on hand gestures only.
This limits their suitability for many applications, such as robot teleoperation, requiring the user to wear a dedicated device or to be very close to the sensors in order for the hands to be detected.
In this work, we propose a gesture recognition system for robot teleoperation that recognises large gestures such as arm waving.
To make the system low-complexity such that it can run online on a mobile robot with limited computing resources, we opt for a modular two-step approach: 1) body pose estimation and 2) gesture classification.



\textbf{Body pose estimation} has been widely studied recently and is used for many applications such as HRI, motion analysis, behaviour prediction, and virtual reality \cite{zheng2020deep}. 
It is often performed using RGB and RGB-D data \cite{toshev2014deep, cao2019openpose}.
In \cite{furst2020hperl}, the authors propose a multi-modal system using RGB imagary and lidar scans to obtain a precise 3D pose estimation.
However, these methods require large CNNs and large amounts of data to achieve good results, since they use high-dimension inputs such as high-resolution images and dense point clouds to predict body poses. This makes them ill-suited for real-time usage with limited computing.
In this work, we pre-process lidar scans to preserve only the relevant information in a low resolution 2D depth image, and use the latter for pose estimation, highly reducing input dimensionality.

\textbf{Gesture classification} can be achieved using end-to-end models with video signals as inputs \cite{miao2017multimodal, zhu2017multimodal, zhou2021regional}. In \cite{zhou2021regional}, the authors use two separate 3D CNNs to learn spatio-temporal features of RGB and and depth sequences. A Convolutional LSTM is used to learn the long-term spatio-temporal features for each separate input stream. Finally, the two streams are fused by simple averaging to obtain the final result.
Others also leverage body pose estimation systems to build gesture recognition modules from skeleton data \cite{asadi2017survey}. 
Fully connected layers together with temporal matching mechanisms are used to detect gestures from skeleton data \cite{li2019skeleton}.
Graph Convolutional Networks can also recognize gestures \cite{yan2018spatial}, and predict future body poses from the input sequences \cite{li2019actional}.  
In \cite{Vaughan1}, the authors use a CNN to detect a simple body pose (hands and face) in UAV video frames. The relative position of the hands to face is used to detect the gestures and execute a robot task. The authors extend this work in \cite{Vaughan2} to also compute a 3D pointing vector for an RGB-D camera and use its intersection with the floor as a waypoint for the UGV to manoeuvre to. 
Other work focuses on the use of Recurrent Neural Networks (RNNs) for the purpose of gesture recognition (e.g. DeepGRU is a network based on multiple stacked GRU units that performs gesture recognition from raw skeleton, pose or vector data \cite{maghoumi2019deepgru}).
We opt for this approach of using only pose data to predict gestures with a RNN because of the low dimensionality of this type of data and the RNN's capacity to process temporal information. 
Finally, RGB frames, 3D skeleton joint information, and body part segmentation can be combined using a set of neural networks, such as a 3D CNN and a LSTM network, to predict human gestures \cite{nguyen2020gesture}.



\section{Gesture Recognition System}

\begin{figure*}[t]
  \includegraphics[width=\textwidth]{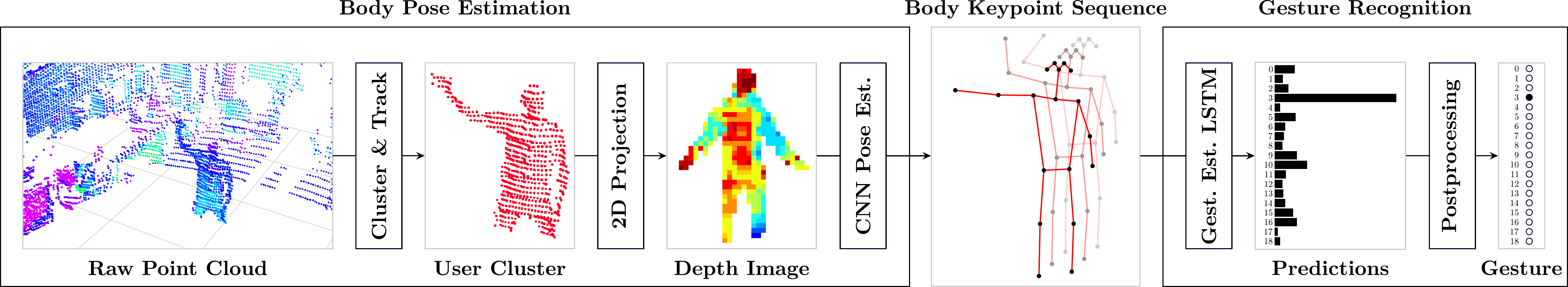}
  \caption{Our modular gesture recognition system. The body pose estimation module takes lidar data as inputs. The person in each point cloud frame is segmented and projected as a 2D depth image before being used by the CNN to predict a body pose in the form of keypoint coordinates. Then, keypoints are accumulated to form a sequence, which is fed to the gesture classifier at every iteration to predict human gestures. }
  \label{fig:overview}
\end{figure*}

We propose a robust and modular gesture recognition system using lidar as input.
Figure \ref{fig:overview} provides an overview of the system's architecture.
The system consists of two main components: a body pose estimator and a gesture classifier.
On the left, we see the body pose estimation system, which uses a three step procedure to predict the user's body pose from a lidar scan.
On the right, we see that the gesture classification module uses a sequence of the estimated poses to predict user gestures.

This modular approach has many advantages: 1) the predicted body poses act as an intermediate explainable representation; 2) the gesture classifier is independent of the modality used to predict keypoints; 3) the pose estimation system reduces significantly the input dimensionality, which makes the gesture classifier lightweight.

\subsection{Pose Estimation Network}\label{sec:pose-estimation}

The proposed pose estimation system consists of three main modules: 1) 3D segmentation, 2) 2D projection and 3) pose estimation.
The first module pre-processes the lidar point cloud to segment the points that belong to the tracked person.
A cluster tracking algorithm can continuously track a person 360 degrees around the sensor to recognise his gestures.
The algorithm uses euclidean clustering to extract clusters from the point cloud and a Kalman filter to track the user given an initial position \cite{kalman}. Extensions to multi-person tracking to enable multi-user UGV control are currently being investigated.
Omnidirectional tracking is essential for vehicle teleoperation, as both the vehicle and the user can move.
The user cluster point cloud is projected on a plane and the module generates a one channel $128 \times 64$ pixel depth image from the sensor's perspective.
This reduces the data complexity while keeping all relevant information about the person's pose.
Finally, this image is fed to the pose estimation CNN, which estimates 8 body keypoint positions: the hips, the shoulders, the elbows and the wrists.
The proposed network is composed of four convolutional layers with $3 \times 3$ kernels, 64 channels and a stride of one followed by two fully connected layers, of size 512 and 256 respectively. 
Each convolutional layer is followed by a $2 \times 2$ kernel maxpool.
All activation functions are ReLUs, except the last one, which is a sigmoid.

\subsection{Gesture Classifier}

We frame the gesture classification problem as a temporal problem and use a network based on a LSTM to classify a sequence of estimated body poses. 
The sequence contains approximately one second of data, which corresponds to between 10 and 30 frames, depending on the frame rate of the input data (e.g. 10Hz for a lidar, and 30Hz for a stereo camera).
The network is composed of one LSTM layer with 50 hidden dimensions followed by a fully connected layer and a softmax.
When using the system for teleoperation, an extra filtering step of post-processing is added to the system's predicted gestures.
For this purpose, a buffer of predicted gestures is created.
A certain threshold (gesture count of the same type within the buffer window) needs to be achieved to trigger a gesture prediction, and a lower threshold needs to be maintained to continue predicting the current gesture.
This makes the changes in gestures smooth and avoids high frequency glitches for both static and dynamic gestures.


\section{Methodology}

The proposed system is low-complexity and only requires a small dataset with weak labels, both for pose estimation and for gesture classification.
In fact, we propose an automated labeling setup for pose estimation from the lidar using the capabilities of an existing pose estimation system from stereo imagery.
We also introduce multiple data augmentation strategies to improve the classifier and make it independent of the body pose estimation method.

\subsection{Training for Lidar Pose Estimation}

To train our pose estimation network, we leverage a pose estimation system from stereo imagery. 
We collected synchronised lidar scans and 3D body pose predictions and we then follow the procedure described in section \ref{sec:pose-estimation} to obtain training depth images with no labeling efforts required.
We also augmented the samples with different transformations such as rotations, translations and resizing in order to generate more samples. 
In total, we collected 3960 frames for training over a period of 20 minutes from two different individuals, which amounts to 738000 samples after the augmentation process. 
Our CNN was trained over 20 epochs, with a learning rate of 0.001.
A total of 1000 frames were collected for testing, from three individuals different from the two used for training.
For our experiments, we used the Ouster OS1-64 lidar and the ZED2 stereo camera along with its Software Development Kit (SDK), which offers a working pose estimation module\footnote{https://www.stereolabs.com/docs/}.



\subsection{Sample-efficient Gesture Classification Training}

\begin{figure}[h]
  \centering
  \includegraphics[width=\linewidth]{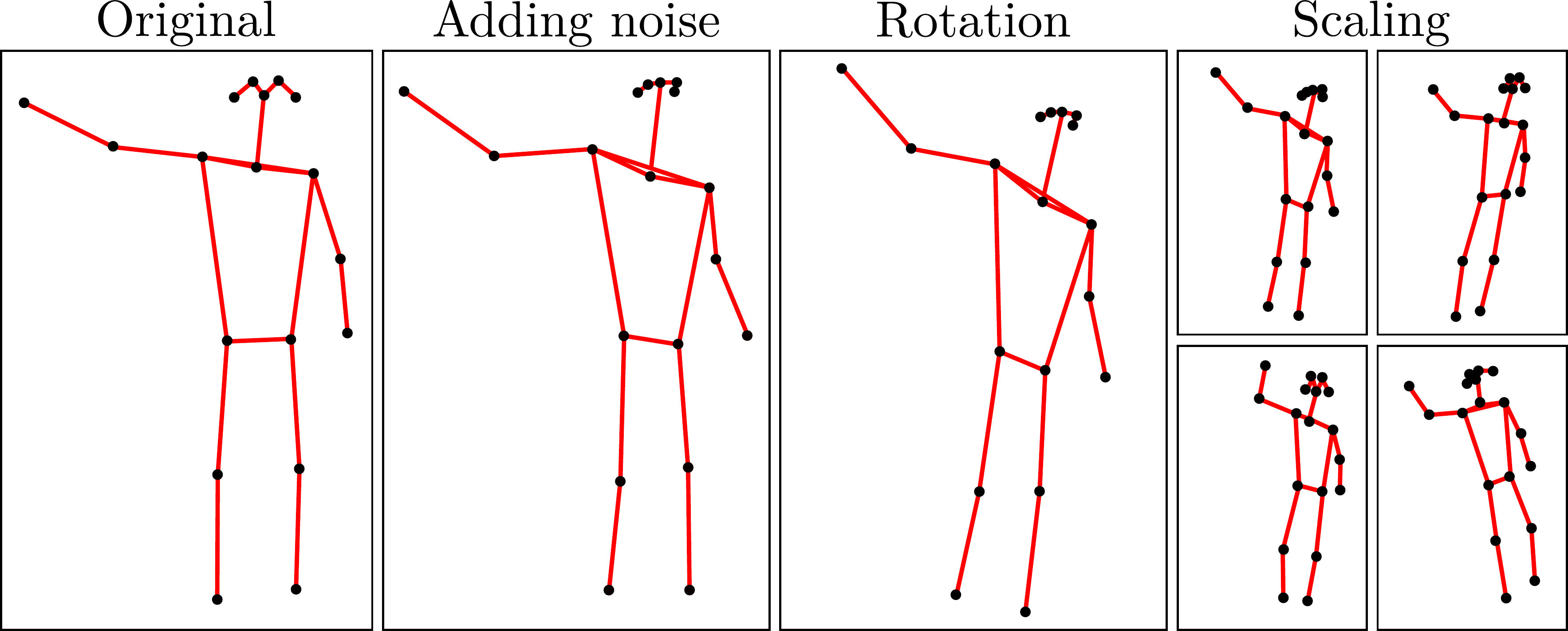}
  \caption{Data augmentation pipeline for gesture classification training. First, a sequence is sampled from the original recording using a varying time scaling factor. Then, constant random noise is added to the body keypoints for all the sequence. Finally, transformations such as rotations, translations and size scaling are applied.}
  \label{fig:augmentation}
\end{figure}

To train our gesture classifier, we collected body pose data only from stereo input.
We collected gestures from eight individuals, each doing one sample of each of the 18 learned gestures for approximately six seconds and several negative samples containing no gesture of interest.
The data from four people was used for training and the rest was used as the test set.
To generate the training dataset, we sampled one second sequences from each six second recording and applied multiple transformations to the body keypoints including noise, resizing, rotation and translations.
The time scale was also altered by dropping frames or adding intermediate frames using interpolation.
The augmentation process is illustrated in figure \ref{fig:augmentation}.
This data augmentation process allows us to train a useful gesture classifier with only 80 original gesture recordings.
As mentioned, no extra data collection is needed to train a classifier for experiments with the lidar. 
We use the same data for training, but alter the augmentation process by simulating a 10Hz frame rate, which is similar to the lidar's frequency of acquisition.
Both of our LSTM-based gesture classifiers have the same architecture and are trained using a learning rate of 0.001.

\section{Experimental Results}
The performance of each subsystem presented in the previous section is evaluated individually. 
Then, the full pipeline is tested in terms of maximal range and maximum body rotation with respect to the sensor.
We also test the responsiveness of our system to determine the average delay to predict a certain gesture.
The ease of use of our teleoperation system is also qualitatively evaluated.

\subsection{Pose Estimation Accuracy}

Table \ref{tab:keypoint-results} shows the mean error for each body keypoint from the test set.
Since the network is trained by using the keypoints predicted by the stereo pose estimation module as labels, the accuracy is measured by using them as ground truth values even though they are not perfectly accurate.
The results show that our system has an average error of 4.3 cm when compared to the stereo pose estimation system used for labeling. 
We can see that the hips and the shoulder features have a lower error, and the elbow and wrist features have a higher error, with an average of 5.6 cm.
Compared to stereo imagery, the lidar is appealing as it has a higher range of operation for pose estimation (up to 10m), is robust to any lighting condition, and works 360 degrees around the sensor.
These advantages are critical for the application of the system to vehicle teleoperation, since it gives more freedom to the user.
Furthermore, because we leverage an existing stereo pose estimation system, no manual labeling was needed for training.

\begin{table}[h]
\centering
\caption{Pose estimation mean ($\mu$) and standard deviation ($\sigma$) error (in cm) for right (R) and left (L) body keypoints. }
\begin{tabular}{c|cc|cc|cc|cc|c}
    \toprule
    & \multicolumn{2}{c|}{\textbf{Hips}} & \multicolumn{2}{c|}{\textbf{Shoulders}} & \multicolumn{2}{c|}{\textbf{Elbows}} & \multicolumn{2}{c|}{\textbf{Wrist}} & \multirow{2}{*}{\textbf{All}} \\
    & \textbf{R} & \textbf{L} & \textbf{R} & \textbf{L} & \textbf{R} & \textbf{L} & \textbf{R} & \textbf{L} \\
    \midrule
    $\mu$ & 3.1  & 3.0  & 2.9  & 3.0  & 6.3  & 4.4 & 5.5  & 6.3 & 4.3  \\
    $\sigma$ & 0.2  & 0.2  & 0.2  & 0.2  & 0.3  & 0.3 & 0.4 & 0.4 & 0.3  \\
    \bottomrule
\end{tabular}
\label{tab:keypoint-results}
\end{table}

\subsection{Gesture Classifier Performance}

\begin{table*}[h]
\centering
\caption{Results of the proposed gesture classifier on test set.}
\begin{tabular}{cc|cccccccccccc|cccccc}
\toprule
& & \multicolumn{12}{c}{\textbf{Static}} & \multicolumn{6}{|c}{\textbf{Dynamic}}    \\
& \textbf{Gestures} & \textbf{1} & \textbf{2} & \textbf{3} & \textbf{4} & \textbf{5} & \textbf{6} & \textbf{7} & \textbf{8} & \textbf{9} & \textbf{10} & \textbf{11} & \textbf{12} & \textbf{13} & \textbf{14} & \textbf{15} & \textbf{16} & \textbf{17} & \textbf{18} \\
\midrule
\multirow{2}{*}{Stereo} & Precision (\%) & 100  & 98  & 100  & 94  & 100  & 100 & 96 & 97 & 97 & 100 & 100 & 78 & 98 & 98  & 98  & 96  & 100  & 100  \\
& Recall (\%) & 100  & 100  & 92  &  100 & 87  & 98 & 94 & 100 & 98 & 98 & 100 & 97 & 93  & 96  & 100  & 93  & 85  & 67  \\
\midrule
\multirow{2}{*}{Lidar} & Precision (\%) & 94  & 99  & 95  & 91  & 94  & 95 & 90 & 95 & 92 & 87 & 94 & 94 & 84 & 94  & 95  & 84  & 99  & 98  \\
& Recall (\%) & 77  & 91  & 91  & 74  & 91  & 92 & 92 & 92 & 80 & 87 & 92 & 90 & 66  & 76  & 78  & 84  & 71  & 55  \\
\bottomrule
\end{tabular}
\label{tab:gesture-results}
\end{table*}

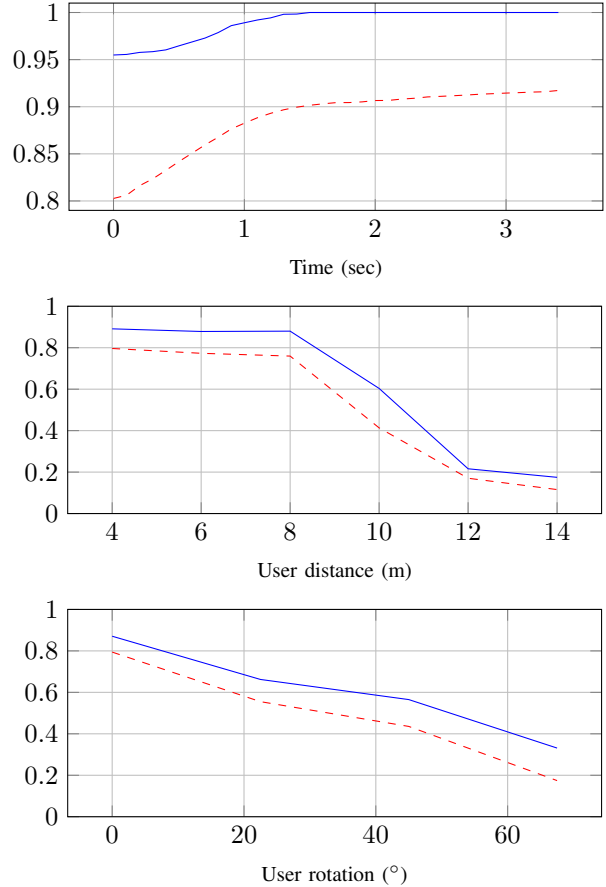
\begin{figure}
    \centering
    \begin{subfigure}[b]{\linewidth}
        \begin{tikzpicture}
        \begin{axis}[height=0.5\linewidth, width=\linewidth, xlabel=\footnotesize Time (sec), ymin=0.79, ymax=1.01, grid=both]
        \addplot[color=blue] table [x=Time, y=Precision, col sep=comma] {csv/fig4a.csv};
        \addplot[color=red, dashed] table [x=Time, y=Recall, col sep=comma] {csv/fig4a.csv};
        \end{axis}
        \end{tikzpicture}
        \vspace{-5pt}
    \end{subfigure}
    \par\medskip
    \begin{subfigure}[b]{\linewidth}
        \hspace{1pt}
        \begin{tikzpicture}
        \begin{axis}[height=0.5\linewidth, width=\linewidth, xlabel=\footnotesize User distance (m), ymin=0, ymax=1, grid=both]
        \addplot[color=blue] table [x=Distance, y=Precision, col sep=comma] {csv/fig4b.csv};
        \addplot[color=red, dashed] table [x=Distance, y=Recall, col sep=comma] {csv/fig4b.csv};
        \end{axis}
        \end{tikzpicture}
        \vspace{-5pt}
    \end{subfigure}
    \par\medskip
    \begin{subfigure}[b]{\linewidth}
        \hspace{1pt}
        \begin{tikzpicture}
        \begin{axis}[height=0.5\linewidth, width=\linewidth, xlabel=\footnotesize User rotation ($^\circ$), ymin=0, ymax=1, grid=both]
        \addplot[color=blue] table [x=Angle, y=Precision, col sep=comma] {csv/fig4c.csv};
        \addplot[color=red, dashed] table [x=Angle, y=Recall, col sep=comma] {csv/fig4c.csv};
        \end{axis}
        \end{tikzpicture}
        \vspace{-5pt}
    \end{subfigure}     
    \caption{Gesture recognition system performance (Precision (\textcolor{blue}{--}) and Recall (\textcolor{red}{- -})) with the lidar pose estimation and the gesture classifier for different a) context times for gesture buffering, b) user distances from the lidar and c) body rotations with respect to the lidar. The results represent the mean results of all 18 learned gestures.}
    \label{fig:precision_recall}
\end{figure}


Table \ref{tab:gesture-results} presents the results of our gesture classifier on the test set with no post-processing being applied.
The results show that the model has an average precision of 94.6\% for stereo input and 82.7\% for lidar input.
Noticeably, the dynamic gestures are harder to recognize and have a lower recall average compared to the static gestures, both for stereo and lidar.
On the other hand, the classifier performs better when using body keypoints derived from stereo imagery.
However, for vehicle teleoperation, the use of lidar is crucial for robustness to lighting conditions and omnidirectional usage.
As it is shown in figure \ref{fig:precision_recall}, our post-processing of the network's predictions highly improves performance. 


\subsection{Complete System Performance}

The system was tested on an Argo J8 vehicle\footnote{https://www.argo-xtr.com/index.php/xtr-robots/j8-atlas-xtr/} with  both the body pose estimation pipeline and gesture recognition runing on an Asus Zenbook Laptop with a i7-8565U CPU and a GTX 1050 graphics card.
The system, implemented using PyTorch \footnote{https://pytorch.org/}, is lightweight, as it only requires 500 MB of memory, and ran on CPU.
Inference took less than 50 ms with this setup, including lidar scan pre-processing and gesture prediction post-processing.

Figure \ref{fig:precision_recall} shows the performance of the system, including pose estimation from lidar and gesture classification from the predicted keypoints.
For this experiment, we evaluate performance on 10 second recordings of gestures, where the user has no feedback on the system predictions that would allow him to correct his movements.
We show that the system reaches a perfect precision and a 90\% recall after approximately 1.1 seconds of performing a given gesture.
In the same figure, we evaluate the limits of the system in terms of user distance and rotation. 
Precision and recall are excellent up to eight meters, and the system remains usable up to around ten meters.
The main reason explaining the rapid performance drop upwards of eight meters is the 2D projection step described in section \ref{sec:pose-estimation}.
In fact, the user cluster becomes sparser as the person is further and so does the 2D projection.
A future improvement for the system could be to develop a more robust filtering procedure to produce a better 2D projection from sparse user point cloud clusters.

Figure \ref{fig:precision_recall} also shows the performances with respect to user rotation (rotation of the torso with respect to the lidar centroid).
We notice that the system can perform reasonably well up to a 45 degree rotation, maintaining a 56.5\% precision and a 43.6\% recall.
This performance is sufficient for the system to be usable, especially when the user has access to feedback from the system and can correct the way of performing each gesture depending on the predictions.
This limitation is due to the fact that the lidar scans are projected in 2D, which makes it difficult to classify gestures when the body rotates, but can be easily mitigated by the user facing the vehicle when performing gestures.

\begin{figure}
    \centering
    \includegraphics[width=0.8\linewidth]{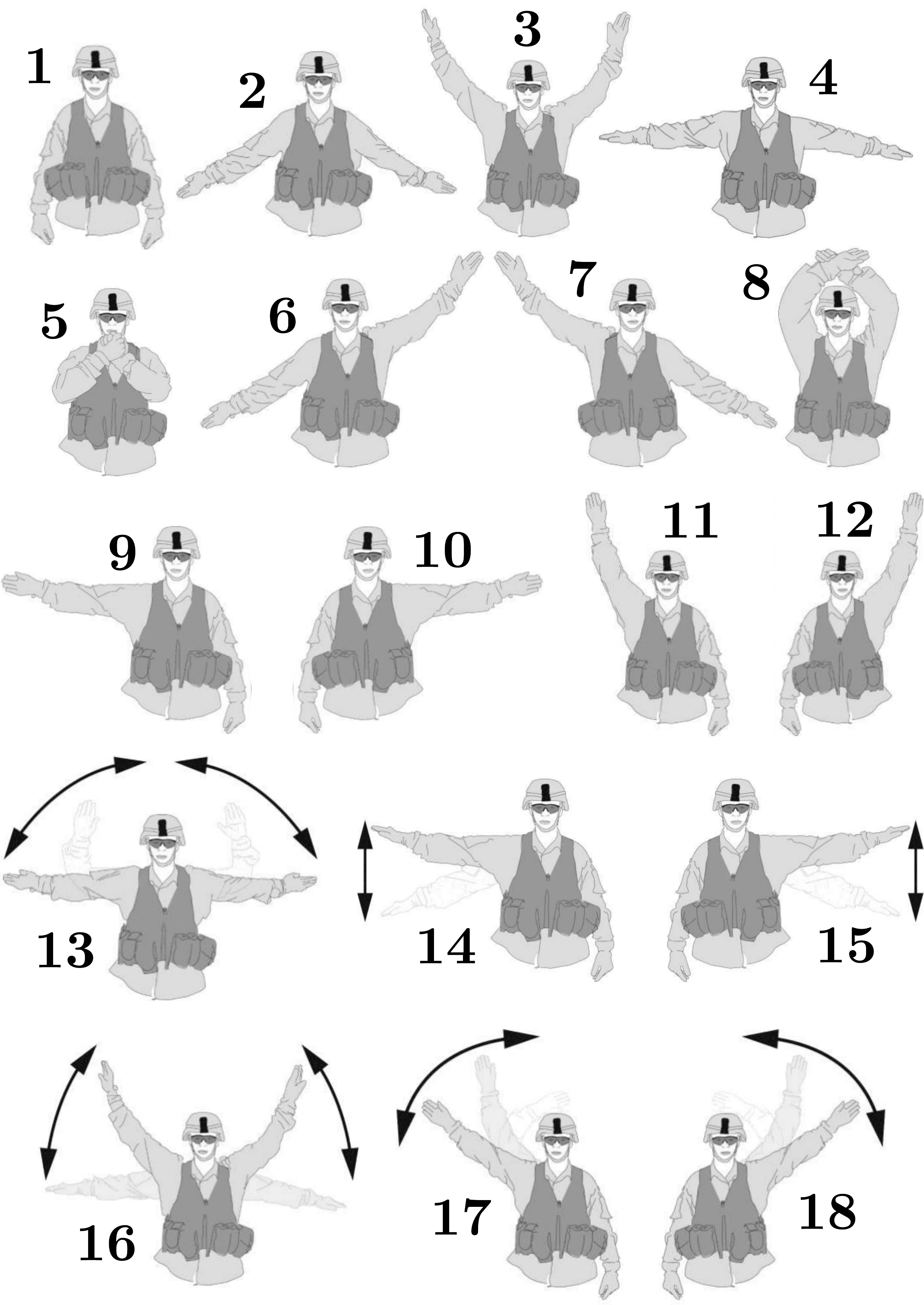}
    \caption{Static (1--12) and dynamic (13--18) gestures learned by the classifier (Images from \cite{VisualSi41:online}). }
    \label{fig:gestures-example}
\end{figure}

\section{Gesture-based Vehicle Teleoperation}

\begin{figure}
    \centering
    \includegraphics[width=\linewidth]{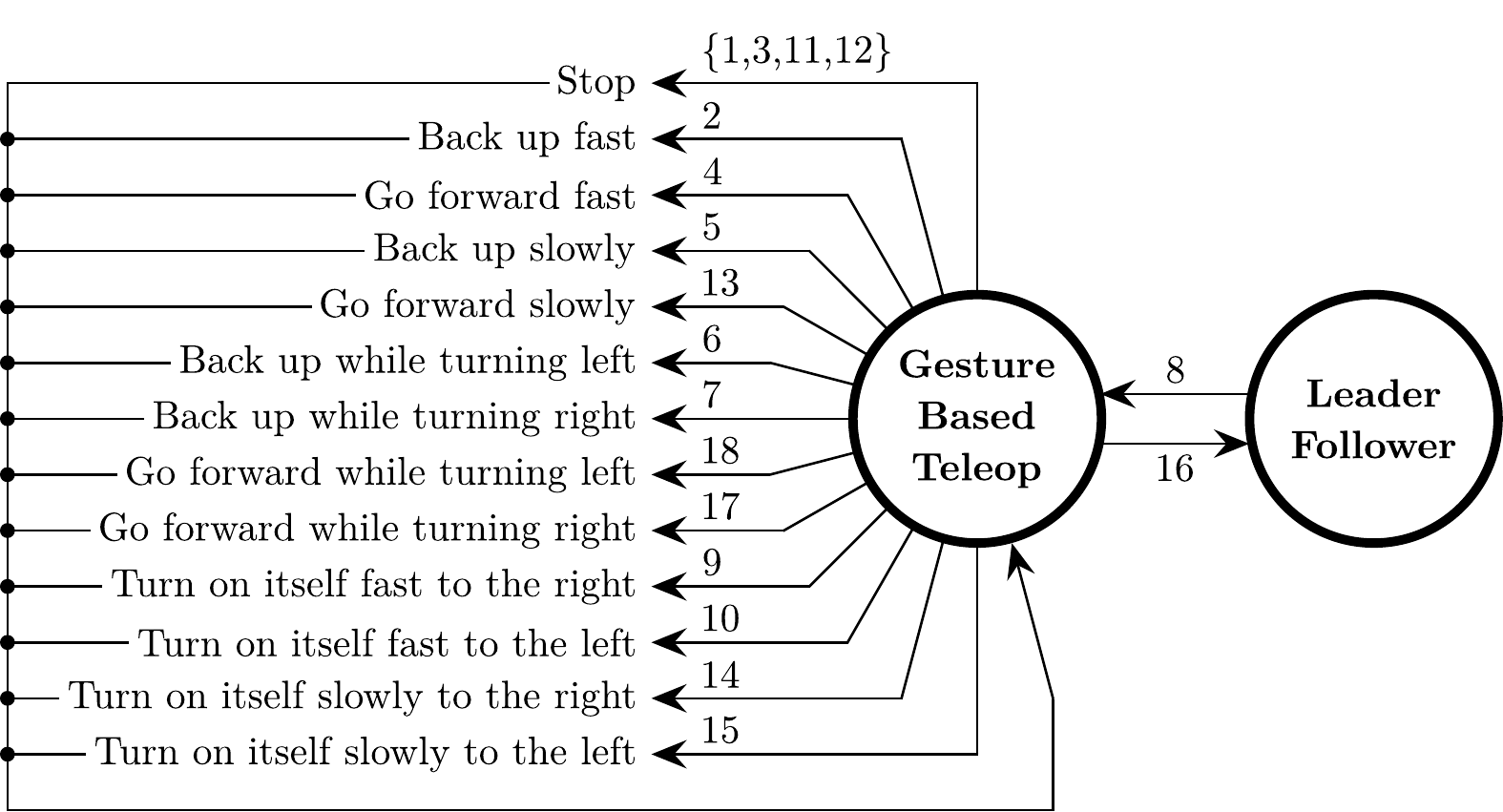}
    \caption{Transitions between states for each gesture (numbered).}
    \label{fig:states}
\end{figure}

We propose an intuitive and complete gesture teleoperation system to showcase the robustness of our gesture recognition system.
There are 18 gestures learned by our system, illustrated in fig. \ref{fig:gestures-example}, and each one of them is mapped to a specific command.
These gestures are inspired by the United States Army visual signals catalog \cite{VisualSi41:online}, but the proposed teleoperation system maps them to different commands.
Figure \ref{fig:states} shows the different state transitions triggered by the learned gestures.
The system supports two modes: 1) a gesture-based teleoperation mode and 2) a leader-follower mode. 
When in gesture-based teleoperation mode, the vehicles executes commands based on recognised gestures.
When in leader-follower mode, the vehicle simply follows the user until the mode is deactivated.
A cluster tracking algorithm and a proportional controller are used for this purpose.
This mode eases operation when precise control is unnecessary and is more suitable for traveling longer distances.
Specific gestures are chosen to enable and disable the leader-follower mode.

\section{Conclusion}

We propose a novel, robust and low-complexity lidar-based gesture recognition system for vehicle teleoperation.
We present a pose estimation system that uses a CNN to track and extract user body keypoints. These keypoints are input to the proposed gesture classification module that uses a LSTM network and a temporal post-processing procedure to classify gestures. Finally, a working teleoperation setup is designed to showcase system teleoperation from lidar input.
The system can run in real-time on a robotic vehicle with limited computing capabilities, requires very little data collection and no manual labeling of gestures.
The system was able to function in all tested outdoor conditions, regardless of lighting and the users position around the vehicle.
The results show a lot of potential for gesture-based robotic control.

Future development will focus on a multi-user extension allowing for control hand-off between users as well as the integration of the system as part of a multi-modal framework for robot teleoperation that also includes speech and a tablet interface. Finally, this multi-modal system will be used in human factors testing to determine the optimal interface for Canadian Armed Forces operational scenarios.


\bibliography{bibliography}
\bibliographystyle{IEEEtran}

\end{document}